\documentclass[aps,prb,showpacs,superscriptaddress,amsmath,twocolumn]{revtex4}

\usepackage{amsmath}
\usepackage{amssymb}
\usepackage{graphicx}

\begin{document}

\title{Semiclassical study of edge states and transverse electron focusing for strong spin-orbit coupling}

\author{Andor Korm\'anyos}
\thanks{e-mail: a.kormanyos@lancaster.ac.uk}
\affiliation{Department of Physics, Lancaster University,
Lancaster, LA1 4YB, United Kingdom}

\begin{abstract}
 We studied the edge states and transverse electron focusing in the presence of spin-orbit interaction 
 in a two dimensional electron gas. Assuming strong spin-orbit coupling we derived semiclassical 
 quantization conditions to describe the  dispersion of the edge states.  
 Using the disprsion relation we then 
 make  predictions about certain  properties of the focusing spectrum. 
 Comparison of our analytical results with quantum 
 mechanical transport calculations reveals that  certain features of the focusing spectrum can be 
 quite well understood in terms of the interference of the edge states while the explanation of other 
 features seems to require a different approach.  
\end{abstract}


\maketitle

\section{Introduction}

\begin{figure}[tb]
\includegraphics[scale=0.45]{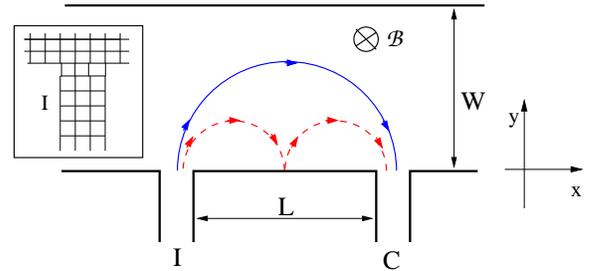}
\caption{(color online)  Schematic geometry  of the transverse electron focusing
setup. The 2DEG is contacted by an injector (I) and a collector (C) 
probe and  perpendicular magnetic field  is applied. 
Classical quasiparticle trajectories leaving from the injector at normal
direction, depending on the strenght of the magnetic field, can  be focused 
onto the collector.  The inset shows the details of the tight-binding model 
used to describe the 
the probes (c.f. the inset of Fig.~1 in Ref.~\onlinecite{usaj}). 
\label{fig:focusing-geometry}}
\end{figure}

Semiclassical approximations  are often very practical for understanding certain
physical phenomena. Beyond their practicality,  they also provide a rather general framework 
to treat quantum systems of interest. 
A  good example to illustrate the merits of semiclassical approach is  the 
transverse electron focusing (TEF). 
The  geometry of  electron focusing is shown in Fig.\ref{fig:focusing-geometry}. 
The current is injected into the 
sample at  a quantum point contact called injector (I) in perpendicular magnetic
field $\mathcal{B}$.  
If the magnetic field is an integer multiple of a focusing field $B_{focus}$, 
electrons injected within a small angle around the perpendicular direction to
the edge of the sample can be focused onto a collector  quantum
point contact (denoted by $C$ in Fig.~\ref{fig:focusing-geometry})  
which acts as a voltage probe. 
Therefore, if  the collector voltage is plotted
as a function of  magnetic field one can observe equidistant peaks at magnetic fields
$B=p*B_{focus}$ ($p=1,2,3,\dots$) corresponding to cases where 
the cyclotron diameter $2 R_c(\mathcal{B})$  is an integer multiple of the distance $L$ between
the injector and the collector.

TEF is a versatile experimental technique (for a review of the various problems 
where it has been used see Ref.~\onlinecite{tsoi}). 
In the case of quantum wells containing two dimensional electron gas (2DEG), 
the accessibility of  the quantum ballistic transport regime  opened up the way to the experimental 
demonstration of \emph{coherent electron focusing}\cite{Houten_Carlo,lu}  as well. 
Recently, several experiments investigated the effect of spin-split bands in semiconductors
on magnetic focusing\cite{potok,ref:rokhinson,dedigama}. Of special interest are for us the 
experiments of Refs.~\onlinecite{ref:rokhinson,dedigama} 
in which evidence of spin-orbit interaction (SOI) dependent focusing have been found. These
experiments sparked  considerable theoretical interest\cite{usaj,reynoso-2,zulicke,schliemann} 
as well. Refs.~\onlinecite{usaj,reynoso-2,zulicke,schliemann} have in common
that they consider the properties of bulk Landau levels in the presence of SOI  
to explain the experimental results on focusing. While using the bulk Landau
levels as a starting point is certainly justified when discussing 
magneto-oscillations\cite{keppeler-1,lagenbuch}, for the geometry shown 
in Fig.~\ref{fig:focusing-geometry} one  expects that the edge states should 
play a central role in the transport phenomena. Indeed, this was the approach adopted 
in  Ref.~\onlinecite{Houten_Carlo} to discuss coherent electron focusing. 
The rich physics brought about by the interplay of SOI and the confinement due to 
external magnetic 
field and electrostatic potential has also attracted significant theoretical 
attention\cite{wang,pala,reynoso-1,debald,yun-juan,grigoryan} but  implications 
on electron focusing have not been considered. 

Here we aim to investigate whether the electron focusing spectrum in  
2DEG with strong SOI can be explained in terms of  edge states formed  as  a 
combined  effect of  SOI, magnetic field  and (an assumed) hard wall confinement potential.   
To this end we first derive semiclassical quantization conditions which describe the 
dispersion relation of the edge states  in the limit of strong SOI and weak magnetic
fields. These results shed new light on and  help to better understand
 the exact quantum solution of this problem, published very recently 
in Ref.~\onlinecite{grigoryan}. We then study how the properties of the edge states
are manifested in the transport phenomena of the focusing setup  
shown in Fig.~\ref{fig:focusing-geometry}.  We expect that our results should be 
relevant in the case of e.g. InSb quantum wells, where theoretical predictions\cite{gilbertson-2}
and recent experiments\cite{hong,dedigama,khodaparast,gilbertson-1}
 indicate that it is possible to fabricate samples with 
strong (compared to GaAs/AlGaAs heterostructures) spin orbit interaction and 
ballistic quasiparticle propagation over  distances of the order of $1\mu{\rm m}$ at low temperatures. 

The rest of the  paper is organized in the following way: in the next section we briefly introduce the
semiclassical framework that we will be using. In Section~\ref{sec:semiclass} we derive a pair of 
semiclassical quantization conditions for the edge states and compare the obtained 
band structure to the results of
exact numerical calculations. We also discuss how  our semiclassical results are related to
other approximation methods found in the literature. 
The semiclassical  quantization conditions then allows us in Section~\ref{sec:focusing} 
to make prediction about the focusing spectrum. We end our paper by a comparison of these
predictions to numerical transport calculations and a short summary in Section~\ref{sec:summary}.

\section{Semiclassical theory with spin degrees of freedom}
\label{sec:semi-overview}

Generally, the non-relativistic single-particle Hamiltonian of spin $1/2$ particles can be 
written as\cite{amand} 
\begin{equation}
 \hat{H}=\hat{H}_0+\hat{H}_1, \quad \hat{H}_0=\frac{\hat{\mathbf{p}}^2}{2 m^*}+V(\mathbf{r})
\label{eq:gen-so} 
\end{equation}
where  $\mathbf{r}$ and $\hat{\mathbf{p}}$ denotes the position and momentum operators, respectively, 
$m^*$ is the effective mass of the particles and we assume that the spin dependent part 
have the following form: 
\begin{equation}
 \hat{H}_1=\hbar\,\kappa\, \widehat{C}(\mathbf{r},\hat{\mathbf{p}})\cdot\boldsymbol{\sigma}.
\label{eq:gen-so-term}
\end{equation}
Here $\boldsymbol{\sigma}=(\sigma_x,\sigma_y,\sigma_z)$ is a vector of Pauli matrices and the 
$\hbar$ comes from the spin operator $\hat{\mathbf{s}}=\frac{1}{2}\hbar\boldsymbol{\sigma}$.  The constant
$\kappa$ gives the spin orbit coupling. 

The application of semiclassical methods developed for systems that can be described 
by scalar Hamiltonians\cite{brackkonyv} is not straightforward if one has to consider the 
spin degree of freedom as well.  Namely, as a first step 
one would have to define a classical Hamiltonian which is 
not a trivial task since  there is no classical analogue of the spin. Various semiclassical
schemes have been proposed, see Refs. \onlinecite{littlejohn,bolte,pletyukov,keppeler-2}, 
and we refer to these original papers  for most of the details. A short 
overview  of the different approaches can also be found in Refs.~\onlinecite{amand,zulicke}.

Here we will follow the approach first used by Yabana and Horiuchi\cite{yabana} and later 
generalized and further  developed in Refs.~\onlinecite{littlejohn,bolte,keppeler-2,gyorffy,samokhin}. 
It is often called the ``\emph{strong coupling limit}'' because it corresponds to a double limit 
$\hbar\rightarrow 0$ and $\kappa\rightarrow\infty$ while $\bar{\kappa}=\hbar\kappa$ 
is kept constant\cite{amand}. 
For spin $1/2$ particles this approximation scheme leads to two classical Hamiltonians 
\begin{equation}
\mathcal{H}^{\pm}=\mathcal{H}_0\pm\bar{\kappa}|\mathcal{C}(\mathbf{r},\mathbf{p})| 
\end{equation}
where the vector  $\mathcal{C}(\mathbf{r},\mathbf{p})$ is the phase-space symbol of the operator
$\widehat{C}(\mathbf{r},\hat{\mathbf{p}})$\cite{littlejohn,keppeler-2} and represents an 
effective magnetic field which depends on the classical variables $\mathbf{r},\mathbf{p}$.
This approximation  introduces semiclassical phase corrections 
to the orbital motion\cite{yabana,littlejohn}.  
It is restricted however to the case of $|\mathcal{C}(\mathbf{r},\mathbf{p})|>0$ because at phase-space
points where $|\mathcal{C}(\mathbf{r},\mathbf{p})|$ vanishes and therefore $\mathcal{H}^{\pm}$ becomes
degenerate, mode conversion between trajectories described by $\mathcal{H}^{\pm}$ occurs, 
posing a serious difficulty to the theory.

For  a two dimensional electron gas (2DEG) in perpendicular magnetic field and assuming that $\hat{H}_1$ describes 
Rashba-type spin-orbit coupling, which will be our main interest in the 
rest of the paper, the ``strong coupling'' approach results in the following semiclassical 
Landau-level spectrum\cite{amand,reynoso-1}:
\begin{equation}
E_n^s=\hbar\omega_c\left[n \pm\sqrt{2 \,n}\,k_{so}l_B\right], \,\,\,n=1,2,\dots
\end{equation}
where $\omega_c=\frac{e\mathcal{B}}{m^*}$ is the classical cyclotron frequency, 
$l_B=\sqrt{\hbar/e\mathcal{B}}$ is the magnetic length, and by 
using the notation $\alpha_R$ for the coupling constant $\bar{\kappa}$ in this particular case, 
$k_{so}$ is given by $k_{so}=m^*\alpha_R/\hbar$.
Comparing this to  the exact result\cite{ref:bychkov} 
\begin{eqnarray}
E_0 &=&\hbar\omega_c/2\nonumber;\\
E_n &=&\hbar\omega_c\left[n\pm\sqrt{2\,n\,(k_{so}l_B)^2+\frac{1}{4}}\,\right] 
\label{eq:exact-Landau}
\end{eqnarray}
we see that the semiclassical Landau levels  are  good approximations of the 
exact ones if $2 n (k_{so}l_B)^2 \gg 1/4 $. This requires 
large quantum numbers $n$ (i.e. large Fermi energy) 
and/or strong spin-orbit coupling $\alpha_R$ and 
not too strong  magnetic field (i.e. $l_B\sim 1/\sqrt{\mathcal{B}}$ is not too small). 
We expect therefore that the strong coupling method should be adequate 
if these conditions are met. An estimate of the appropriate
 magnetic field range  assuming  InSb quantum well  material parameters will be 
given after Eq.~(\ref{eq:semi-quant-cond}).

\section{Semiclassical theory of edge states}
\label{sec:semiclass}

We assume that the 2DEG, formed e.g. in the quantum well of an InSb heterostructure, 
is in a  perpendicular homogeneous
magnetic field. The motion of electrons is confined by a  hard-wall potential, $V(y)=\infty$ for 
$y<0$ (see Fig.\ref{fig:focusing-geometry} for the geometry). 
The quantum mechanical description of the system can be obtained using the Hamiltonian 
(\ref{eq:gen-so})
where $\hat{H}_0=\frac{\hat{\boldsymbol{\pi}}^2}{2m^*}$ corresponds  to the kinetic energy of particles 
and the operator $\hat{\boldsymbol{\pi}}$ is defined as  
$\hat{\boldsymbol{\pi}}=(\hat{\pi}_x,\hat{\pi}_y)=\hat{\mathbf{p}}+e\mathbf{A}$, where 
$\hat{\mathbf{p}}=-i\hbar\boldsymbol{\nabla}$ is the canonical momentum    operator 
and $\mathbf{A}$ is the vector potential. Furthermore, 
$H_1=\alpha_R(\hat{\pi}_x\sigma_y-\hat{\pi}_y\sigma_x)$ describes the Rashba spin-orbit (RSO) 
coupling in the system, $\sigma_x$, $\sigma_y$ are Pauli matrices acting in the spin space. 
We assume that $\alpha_R$ is constant in space and neglect its possible random variaton due to 
nanosize domains\cite{sherman}.

To preserve the translational invariance of the system, we choose the Landau gauge 
$\mathbf{A}=(\mathcal{B} y, 0, 0)^T$.
Using the ansatz $\Psi(\mathbf{r})=e^{i k x}\Phi(y)$ we can simplify the problem to an effectively
one dimensional (1D) one, which we will solve in semiclassical approximation. The discussion goes along the 
lines of Refs.~\onlinecite{yabana,keppeler-2,gyorffy} [for a recent application 
see also Refs.~\onlinecite{carmiere,sajat2}]. We seek the solutions of the 1D Schr\"odinger equation 
$
\hat{H}\Phi(y) =E \Phi(y)
$ 
in the following form \cite{keppeler-2}:
\begin{equation}
\Phi(y)=\sum_{q\ge 0} \left(\frac{\hbar}{i}\right)^{q} 
 \mathbf{a}_q(y) e^{\frac{i}{\hbar} S(y)},
\label{eq:semiclass_wavef}
\end{equation}
where $\mathbf{a}_q(y)$ are  spinors and 
$ S(y)$ is the classical action. Performing the unitary transformation 
$
\Phi\rightarrow e^{-\frac{i}{\hbar} S(y)}\Phi(y)
$, 
$\hat{H}\rightarrow
e^{-\frac{i}{\hbar} S(y)}\hat{H}\,
e^{\frac{i}{\hbar} S(y)}
$
the Schr\"odinger equation can be rewritten  as 
\begin{equation}
\begin{split}
\left(
\begin{array}{cc}
 \frac{\hat{\Pi}_x^2+\hat{\Pi}_y^2}{2m^*}-{E} & i\alpha_R (\hat{\Pi}_x-i \hat{\Pi}_y)\\
  -i\alpha_R (\hat{\Pi}_x+i \hat{\Pi}_y) &  \frac{\hat{\Pi}_x^2+\hat{\Pi}_y^2}{2m^*}-{E} 
\end{array}
\right)
\left(\mathbf{a}_0(y)\right.+\\
\left.\frac{\hbar}{i}\mathbf{a}_1(y)
+\dots \right)=0.
\label{eq:semiclass_1D_schr}
\end{split}
\end{equation}
Here $\hat{\Pi}_x\equiv\Pi^{0}_x=p_x +e A_x(\mathbf{r})$, $p_x$ being $p_x=\hbar k$, and 
 $\hat{\Pi}_y=\hat{p}_y+\Pi^0_y$, where 
$\Pi^{0}_y=\frac{\partial S(y)}{\partial y}$. The WKB strategy\cite{brackkonyv} 
is to satisfy Eq.~(\ref{eq:semiclass_1D_schr}) separately order by order in $\hbar$. 

At $\mathcal{O}(\hbar^0)$ order we obtain 
\begin{equation}
\left(
\begin{array}{cc}
\frac{(\Pi_x^0)^2+(\Pi_y^0)^2}{2m^*}- {E} & i \alpha_R ({\Pi}_x^0-i {\Pi}_y^0)\\
  -i\alpha_R({\Pi}_x^0+i {\Pi}_y^0) & \frac{(\Pi_x^0)^2+(\Pi_y^0)^2}{2m^*}-{E} 
\end{array}
\right)
\mathbf{a}_0(y)=0.
\label{eq:hbar0ord}
\end{equation}
Nontrivial zeroth order eigenvectors $\mathbf{a}_0(y)$ exist if 
\begin{equation}
E= \frac{Q^2}{2m^*}\pm\alpha_R\, Q 
\label{eq:class_Ham}
\end{equation}
where $Q=\sqrt{(\Pi_x^0)^2+(\Pi_y^0)^2}$. This means that $Q$ is a constant of motion 
for a given energy $E$  and the two branches of Eq.~(\ref{eq:class_Ham}) define
\begin{subequations}
\begin{equation}
 Q_-=\sqrt{p_{so}^2+2 m^* E}-p_{so}
\end{equation}
\begin{equation}
 Q_+=p_{so}+\sqrt{p_{so}^2+2 m^* E}
\end{equation}
\label{eq:defQ}
\end{subequations}
where we used the notation $p_{so}=m^*\alpha_R$. We find therefore that the 
classical equations of motion that can be derived from Eqs.~(\ref{eq:class_Ham}) and (\ref{eq:defQ})  
represent two harmonic oscillators. The corresponding 
zeroth order eigenvectors of Eq.~(\ref{eq:hbar0ord}) are 
\begin{equation}
 \mathbf{V}_{\pm}=\frac{1}{\sqrt{2}}
\left(
\begin{array}{c}
 e^{-\frac{i}{2}(\theta(y)\mp \pi/2)}\\
 e^{\frac{i}{2}(\theta(y)\mp \pi/2)}
\end{array}
\right)
\label{eq:eigenV}
\end{equation}
where $\theta(y)$ is the phase of $\Pi_x^0(y)-i\Pi_y^0(y)$. 
However, the eigenspinor  $\mathbf{a}_0(y)^{\pm}$ can more generally be sought as 
$
\mathbf{a}_0^{\pm}=\mathcal{A}_{\pm}(y) e^{i\gamma_{\pm}(y)}\mathbf{V}_{\pm}
$
where $\mathcal{A}_{\pm}(y)$ is a real amplitude and $\gamma_{\pm}(y)$ is a phase.  
Equations for $\mathcal{A}_{\pm}(y)$  and $\gamma_{\pm}(y)$
can be obtained from the  $\mathcal{O}(\hbar^1)$ order of Eq.~(\ref{eq:semiclass_1D_schr}). 
Using Eq.~(\ref{eq:eigenV}) we find that the $\mathcal{O}(\hbar^1)$ 
order equation can be cast into the following form:
\begin{equation}
\overrightarrow{\nabla}\left(\frac{\mathcal{A}_{\pm}^2}{2}
\left[\frac{1}{m^*}\left(
\begin{array}{c}
 \Pi_x^0\\
\Pi_y^0
\end{array}
\right)
\pm \alpha_R \left(
\begin{array}{c}
 \cos\theta(y)\\
\sin\theta(y)
\end{array}
\right)
\right]
\right) =0.
\label{eq:current}
\end{equation}
This equation does not depend on $\gamma_{\pm}$, which means that  with the choice of 
the eigenvectors shown in  Eq.~(\ref{eq:eigenV}), the phase  $\gamma_{\pm}(y)$ is already
determined up to an unimportant constant factor. Moreover, by rewriting  Eq.~(\ref{eq:current}) as 
\begin{equation}
 \overrightarrow{\nabla}\left(\frac{\mathcal{A}_{\pm}^2}{2}
\frac{\partial\mathcal{H}^{\pm}}{\partial\mathbf{p}}\right)=0
\end{equation}
where $\mathcal{H}^{\pm}=Q^2/2m^2\pm\alpha_R\,Q$, it is easy to see that it expresses 
probability current conservation and it can be solved for   $\mathcal{A}_{\pm}$ using
standard methods\cite{brackkonyv}. From these results  one finds that 
$\Phi(y)$ in semiclassical approximation is given by 
\begin{equation}
\Phi^{\pm}(y)= \frac{1}{\sqrt{2 |\Pi_{y}^{0,\pm}|}}
\left(
\begin{array}{c}
 e^{-\frac{i}{2}(\theta^{\pm}(y)\mp \pi/2)}\\
 e^{\frac{i}{2}(\theta^{\pm}(y)\mp \pi/2))}
\end{array}
\right)
 e^{\frac{i}{\hbar}S(y,y_0^{\pm})} 
\label{eq:Phi-semi}
\end{equation}
where
$
|\Pi_{y}^{0,\pm}|=\sqrt{Q_{\pm}^2-(\Pi_x^{0})^2}.
$
However, as the momentum $\Pi_{y}^{0,\pm}$ is a multi-valued function, we need to introduce
 the  index $j=+1, -1$ do distinguish the different branches. 
The corresponding classical actions $S_j^{\pm}(y,y_{0}^{\pm})$ read 
\begin{equation}
S_j^{\pm}(y,y_{0}^{\pm})=j \int_{y_{0}^{\pm}}^{y} \sqrt{Q_{\pm}^2-[\Pi_x^{0}(y')]^2}\,{\rm d}y'
\label{eq:action-def}
\end{equation}
where as usually, we have chosen the classical turning points as the phase reference points 
for the action. Similarly, the phase $\theta^{\pm}(y)=\theta_j^{\pm}(y)$ is  multivalued as well.

We now have to take into account the confinement potential $V(y)$.
The transverse wavefunctions $\Phi(y)$ shown in Eq.~(\ref{eq:Phi-semi}) would not satisfy the 
boundary condition at $y=0$. In order that the transverse wavefunction does satisfy the boundary condition  
we  make a  linear combination of the functions $\Phi_j^{\pm}(y)$   defined above. 
We try the following ansatz for the transverse semiclassical wavefunction:
\begin{equation}
\begin{split}
 \tilde{\Phi}(y)=&\frac{\mathcal{C}_+}{\sqrt{|\Pi_{y}^{0,+}(y)|}}
\left(
\begin{array}{c}
 e^{i\frac{\pi}{4}}\cos\left(\frac{S_0^+(y,y_0^+)}{\hbar}-\frac{\theta^+(y)}{2}+\frac{\pi}{4}\right)\\
 e^{-i\frac{\pi}{4}}\cos\left(\frac{S_0^+(y,y_0^+)}{\hbar}+\frac{\theta^+(y)}{2}+\frac{\pi}{4}\right)
\end{array}
\right)+\\
&\frac{\mathcal{C}_-}{\sqrt{|\Pi_{y}^{0,-}(y)|}}
\left(
\begin{array}{c}
 e^{-i\frac{\pi}{4}}\cos\left(\frac{S_0^-(y,y_0^-)}{\hbar}-\frac{\theta^-(y)}{2}+\frac{\pi}{4}\right)\\
 e^{+i\frac{\pi}{4}}\cos\left(\frac{S_0^-(y,y_0^-)}{\hbar}+\frac{\theta^-(y)}{2}+\frac{\pi}{4}\right)
\end{array}
\right)
\end{split}
\label{eq:semi-wavefunc-spinor}
\end{equation}
where $\mathcal{C}_+$, $\mathcal{C}_-$ are constants and 
the 
$\pi/4$ factor in the argument of the cosine functions 
takes into account the effect of the classical turning 
 points at $y_{0}^{\pm}$ which appear due to the  magnetic field.  
 The turning points are given by  the physically acceptable zeros 
of the  equation $Q_{\pm}^2-[\Pi_x^{0}(y')]^2=0$. 
Note that  $\tilde{\Phi}(y)$ also 
depends on $k$ through $S_0^{\pm}$ and $\theta^{\pm}$ but in order to keep the notations
 uncluttered  we did not write this out explicitly. 
The dispersion relation for the edge states can be 
obtained by demanding that the wave function vanishes at the boundary, i.e. $\tilde{\Phi}(y=0)=0$.
This is a homogeneous system of equations and nontrivial solutions can only be found if 
the respective determinant is zero. This results in the following implicit dispersion relation:
\begin{equation}
\begin{array}{c}
\cos\left(\frac{S_0^+(0,y_0^+)}{\hbar}-\frac{\theta^+(0)}{2}+\frac{\pi}{4}\right)
\cos\left(\frac{S_0^-(0,y_0^-)}{\hbar}+\frac{\theta^-(0)}{2}+\frac{\pi}{4}\right)+\\
\cos\left(\frac{S_0^-(0,y_0^-)}{\hbar}-\frac{\theta^-(0)}{2}+\frac{\pi}{4}\right)
\cos\left(\frac{S_0^+(0,y_0^+)}{\hbar}+\frac{\theta^+(0)}{2}+\frac{\pi}{4}\right)=0.
\end{array}
\label{eq:implicit-quant}
\end{equation}
Hence, if we denote by $\tilde{\Phi}^{+}(y)$ and $\tilde{\Phi}^{-}(y)$ the first and second
spinors appearing in Eq.~(\ref{eq:semi-wavefunc-spinor}), respectively, the transverse wave function 
can be written, apart from a normalization factor, as  
$
\tilde{\Phi}(y)\sim\tilde{\Phi}^{+}(y)+r_{so}\, \tilde{\Phi}^{-}(y)
$
where $r_{so}$ is given by  
\begin{equation}
r_{so}=-e^{i\frac{\pi}{2}}
\sqrt{\frac{|\Pi_{y}^{0,-}(0)|}{|\Pi_{y}^{0,+}(0)|}}
\frac{\cos\left(\frac{S_0^+(0,y_0^+)}{\hbar}-\frac{\theta^+(0)}{2}+\frac{\pi}{4}\right)}
{\cos\left(\frac{S_0^-(0,y_0^-)}{\hbar}-\frac{\theta^-(0)}{2}+\frac{\pi}{4}\right)}.
\end{equation}
For $k\approx 0$ the reflection amplitude is $r_{so}\approx -e^{i\pi/2}$.   

Furthermore, with the help of trigonometric identities and assuming that 
\begin{equation}
\begin{array}{c}
 \cos\left(\frac{S_0^+(0,y_0^+)}{\hbar}-\frac{S_0^-(0,y_0^-)}{\hbar}\right)\approx
\cos\left(\frac{\theta^+(0)}{2}-\frac{\theta^-(0)}{2}\right)\approx 1
\end{array}
\label{eq:second-ord-diff}
\end{equation}
we find from Eq.~(\ref{eq:implicit-quant}) the following quantization condition:
\begin{equation}
\begin{array}{c}
\cos\left(\frac{\Lambda_+ + \Lambda_- +\varphi_+ +\varphi_-}{2}\right)
\cos\left(\frac{\Lambda_+ + \Lambda_- -\varphi_+ -\varphi_-}{2}\right)=0
\end{array}
\label{eq:quantization-product}
\end{equation}
where for brevity we introduced the notations $\Lambda_{\pm}=\frac{S_0^{\pm}(0,y_0^{\pm})}{\hbar}+\pi/4$ and
$\varphi_{\pm}=\frac{\theta^{\pm}(0)}{2}$. Here we pause for a moment to  interpret 
Eq.~(\ref{eq:second-ord-diff}). From the discussion below Eq.~(\ref{eq:defQ}) and from 
Eq.~(\ref{eq:action-def}) it is clear that if we consider the two branches of Eq.~(\ref{eq:class_Ham}) 
 as classical Hamiltonians for two different quasiparticles then 
 $S_0^{\pm}(0,y_0^{\pm})$ gives (half of the) enclosed flux by the quasiparticle trajectories with the wall
 between two subsequent collisions 
and  $\theta^{\pm}(0)$ is the deflection angle of the momentum between the collisions. 
Therefore  Eq.~(\ref{eq:second-ord-diff}) means   that we neglect the  second and higher
powers of the difference between the enclosed flux and momentum deflection. 

It is clear that Eq.~(\ref{eq:quantization-product}) defines a pair of quantization conditions: 
$\Lambda_+ + \Lambda_- +\varphi_+ +\varphi_-=(2m+1)\pi$ and $\Lambda_+ + \Lambda_- -\varphi_+ -\varphi_-=(2l+1)\pi$
with $l,m=0,1,2,\dots$. By introducing the angle $\beta_{\pm}=\arcsin\left(\frac{X}{R_{\pm}}\right)$, where
$X=k l_B^2$ is the guiding center coordinate and $R_{\pm}=Q_{\pm}/(e\mathcal{B})$ is the radius of the cyclotron
motion for the two quasiparticle branch, we finally arrive at the following two quantization conditions:
\begin{subequations}
\begin{equation}
\begin{split}
\frac{R_+^2}{l_B^2}\left(\frac{1}{2}\sin 2\beta_+ + \beta_+ + \frac{\pi}{2} \right)+
\frac{R_-^2}{l_B^2}\left(\frac{1}{2}\sin 2\beta_- + \beta_- +\frac{\pi}{2}\right)+\\
\left(\beta_+ + \beta_-\right)=4\pi m, \quad m=0,1,\dots m_{max} 
\end{split} 
\label{eq:semi-quant-cond-1}
\end{equation}
and 
\begin{equation}
\begin{split}
\frac{R_+^2}{l_B^2}\left(\frac{1}{2}\sin 2\beta_+ + \beta_+ + \frac{\pi}{2}\right)+
\frac{R_-^2}{l_B^2}\left(\frac{1}{2}\sin 2\beta_- + \beta_- +\frac{\pi}{2}\right)+\\
(\pi-\beta_+) +(\pi-\beta_-)=4\pi l,\quad l=0,1,\dots l_{max}
\end{split}
\label{eq:semi-quant-cond-2} 
\end{equation}
\label{eq:semi-quant-cond}
\end{subequations}
These equations are the first important results of our paper. They do not lend themselves to 
a simple semiclassical interpretation but a possible classical picture could be the 
following: the classical skipping
orbits whose quantization would be described by these equations consist of two segments, each of them 
having slightly different radii given by $R_{\pm}$ but the same guiding center coordinate. 
Besides the orbital motion, the quantization conditions also depend on the change of the phase of 
the spinor part of the wavefunction which is described by the  $(\beta_+ +\beta_-)$, 
$(\pi-\beta_+)+ (\pi-\beta_-)$ terms in Eqs.~(\ref{eq:semi-quant-cond}).  

We note that analogous calculations to the ones outlined above can be carried out if 
the dominant term in the SOI is the $k$-linear Dresselhaus term. Therefore this approach 
can be relevant e.g. in the case of heterostructure studied in Ref.~\onlinecite{akabori}.

\begin{figure}[htb]
\includegraphics[scale=0.8]{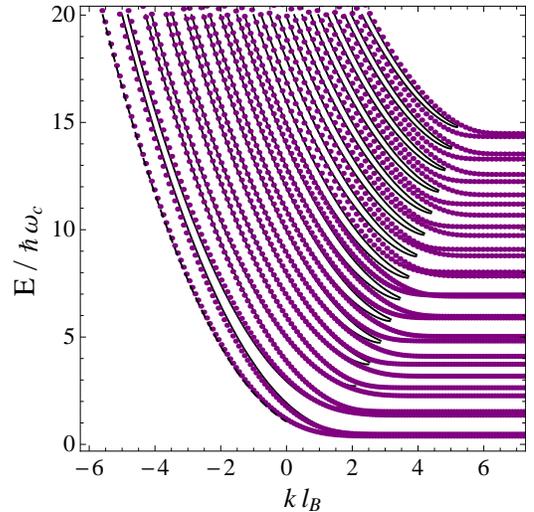}
\caption{(color online) Comparison of the dispersion of the bands as a 
function of the wavenumber $k$ obtained from tight-binding numerical calculations (dots), 
and using the theoretical prediction given by Eqs.~(\ref{eq:semi-quant-cond}) (solid lines) 
for magnetic field  $\mathcal{B}=0.15{\rm T}$ and $k_{so} l_B=0.256$ 
(other parameters are given in the text).
The dashed line at the leftmost band shows the result of Eq.~(\ref{eq:semi-quant-cond-simple})
for $m=0$. 
\label{fig:disp-1}}
\end{figure}
\begin{figure}[htb]
\includegraphics[scale=0.8]{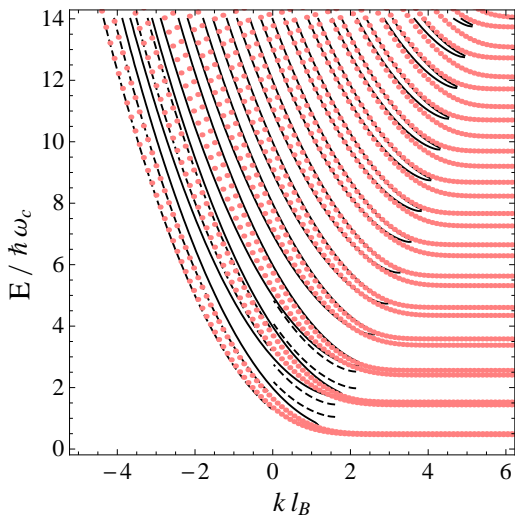}
\caption{(color online) Comparison of the dispersion of the bands as a 
function of the wavenumber $k$ obtained from tight-binding numerical calculations (dots), 
and using the theoretical prediction given by Eqs.~(\ref{eq:semi-quant-cond}) (solid lines) 
for magnetic field $\mathcal{B}=0.6{\rm T}$ and $k_{so} l_B=0.128$ 
(other parameters are given in the text). 
Dashed lines at the leftmost bands show the result of Eqs.~(\ref{eq:semi-quant-cond-simple})
for $m=0,1$. 
\label{fig:disp-2}}
\end{figure}

To see the accuracy of the semiclassical quantization we have 
performed numerical calculations  for the dispersion  of the edge states  
using the tight-binding version of the Hamiltonian (\ref{eq:gen-so}) 
[see e.g. Ref.~\onlinecite{reynoso-2} for the explicit form of the tight-binding Hamiltonian]. 
The results for $k_{so}l_B=0.256$ at a relatively weak magnetic field of $\mathcal{B}=0.15{\rm T}$ 
and using typical parameters
of 2DEG in  InSb quantum well at higher electron densities\cite{gilbertson-2} 
($m^* =0.021\,m_e$, where $m_e$ is the 
bare electron mass and $\alpha_R=1.4*10^{-11}\,{\rm eV m}$) are shown in Fig.~\ref{fig:disp-1}.
As one can see  Eqs.~(\ref{eq:semi-quant-cond}) describe quite well the dispersion of the 
subbands, even at low energies, except for the $k$ values where the guiding 
center  $X\approx R_{\pm}$, i.e. in the transition region to the bulk Landau levels. 
We have found  that although Eqs.~(\ref{eq:semi-quant-cond}) have solutions even for $m=0$ and $l=0$, 
the leftmost band in  Fig.~\ref{fig:disp-1}, which is related to the zeroth Landau level, 
is  poorly approximated by any of the $m=0$ or $l=0$  curves that can be obtained from 
Eq.~(\ref{eq:semi-quant-cond-1}) or   Eq.~(\ref{eq:semi-quant-cond-2}), respectively. Nevertheless,
the approximation works quite well for  rest of the subbands, i.e. for  $m,l\ge 1$. 
For a stronger magnetic field  of $\mathcal{B}=0.6{\rm T}$ ($k_{so}l_B=0.128$) 
shown in Fig.~\ref{fig:disp-2}, 
the approximation for the $m=1$  and $l=1$ bands deteriorated as well, while higher subbands are still
well described by Eq.~(\ref{eq:semi-quant-cond}). 

\begin{figure}
\includegraphics[scale=0.45]{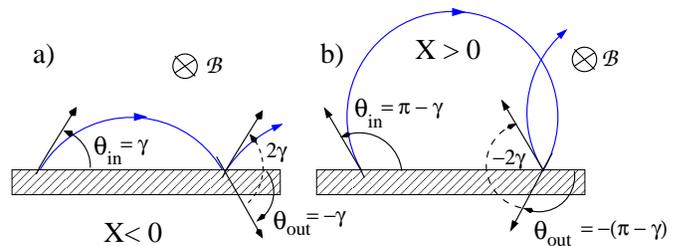}
\caption{(color online) Explanation of the phase shift $-\pi$ for $X>0$ in 
Eq.~(\ref{eq:semi-quant-cond-simple}). Solid (blue) lines show the 
bent quasiparticle trajectories. The direction of the momentum 
at the start of a full period  and at the subsequent collision with the wall 
is indicated by arrows. Also shown are the angles that the momentum encloses with the wall. 
The phase shift contribution
coming from the spinor part of the semiclassical wavefunction is 
$\frac{1}{2}(\theta_{out}-\theta_{in}+2\gamma)$ in (a) and 
 $\frac{1}{2}(\theta_{out}-\theta_{in}-2\gamma)$ in (b).
\label{fig:phase-shift}}
\end{figure}

It is interesting to note that we find that the bands for $0 \ll |X|\lesssim R_{\pm} $ are quite well 
approximated by the following simple formulae: 
\begin{equation}
 \frac{R_{\pm}^2}{l_B^2}\left(\frac{1}{2}\sin 2\beta_{\pm} + \beta_{\pm} + \frac{\pi}{2}\right)-\pi\Theta(X)=
2\pi(m+3/4)
\label{eq:semi-quant-cond-simple}
\end{equation}
where $\Theta(x)=1$ for $x>0$ and zero otherwise. 
Note that in contrast to Eqs.~(\ref{eq:semi-quant-cond})
these equations give a semiclassical quantization of the orbital motion of two independent 
systems whose classical motion is described by the 
Hamiltonians $\mathcal{H}^{\pm}$ given by the left-hand side of Eq.~(\ref{eq:class_Ham}).   
The spinor nature of the quasiparticles enters the quantization only through a 
$-\pi$ phase shift for 
large positive $X$ values. The origin of this phase shift can be understood by looking 
at Fig.~\ref{fig:phase-shift}. 
For $ -R_{\pm} \lesssim X \ll 0$ the phase contribution coming 
from the $\pm\exp(i\theta(y)/2)$ factors of the wavefunction  
(see Eq.~(\ref{eq:eigenV}) is zero over one full period of the classical motion [ Fig.~\ref{fig:phase-shift}(a)]. 
This happens because the phase change accumulated during the orbital motion $\theta_{out}-\theta_{i}=-2\gamma$ is 
canceled by the phase change $2\gamma$ upon reflection when the sign of the $\Pi_y^0$ is negated.  
However, as it is explained  in Fig.~\ref{fig:phase-shift}(b), 
for orbits with $0 \ll X\lesssim R_{\pm}$ the 
total phase change is $-2\pi / 2=-\pi$. We find that the leftmost band in Fig.~\ref{fig:disp-1} which is 
related to the zeroth bulk Landau level (see Eq.~\ref{eq:exact-Landau}) can be quite well 
approximated by the $-$ branch of Eq.~\ref{eq:semi-quant-cond-simple} for $m=0$. For
 stronger magnetic fields, such as  shown in Fig.~\ref{fig:disp-2},  the left-most bands 
are better approximated by the quantization (\ref{eq:semi-quant-cond-simple}) 
than by Eqs.~(\ref{eq:semi-quant-cond}). We note that 
for $X < 0$  the quantization in Eq.~(\ref{eq:semi-quant-cond-simple}) basically 
corresponds to the ``longitudinal SO approximation'' studied in Ref.~\onlinecite{pala}.  

Another interesting and important comparison of the quantization shown in Eqs.~(\ref{eq:semi-quant-cond}) 
can be made to the closely related results of Ref.~\onlinecite{reynoso-2}, where the authors
used a different semiclassical approach\cite{pletyukov} to describe the edge states 
in the presence of RSO. Firstly, we find the same result for the $k=0$ energy 
levels as in Ref.~\onlinecite{reynoso-2}:
\begin{equation}
 E_m(k=0)=\hbar\omega_c(n-(k_{so}l_B)^2)
\end{equation}
and a comparison with the numerical calculations show that it is a good approximation of the 
exact results. However,  we  obtained a pair of quantization conditions, 
not just one [see Eq.~(27) in Ref.~\onlinecite{reynoso-2}].  Furthermore, 
for the parameter range we consider our results 
give a good approximation of the numerically calculated bands not only close to $k=0$ but 
for the whole dispersion relation. Finally, an important difference in the semiclassical  interpretation 
of the skipping orbits is the following: 
in the classical picture put forward in  Ref.~\onlinecite{reynoso-2} 
the skipping orbits consist of two different type of arcs, having radii $R_{\pm} $  and guiding
center coordinates $X_{\pm}$. Moreover, the guiding center changes upon each reflection at the wall 
(see Fig.6 in Ref.~\onlinecite{reynoso-2}). In contrast, our approach tells that the 
guiding center remains the same throughout the motion. We think that this is physically plausible
because the guiding center is a constant of motion. It is instead the reflection angle that 
slightly changes at each collision with the wall as a consequence of having two 
Fermi surfaces with different radii.

\section{Magnetic focusing}
\label{sec:focusing}
Having obtained the quantization condition for edge states in
Eqs.~(\ref{eq:semi-quant-cond}), the calculation of the focusing magnetic fields
$B_{focus}$ goes along the lines of the discussion of Ref.~\onlinecite{Houten_Carlo}. 
We expect that the ballistic transport in a mesoscopic wire in the magnetic field regime
 where the cyclotron diameter 
is smaller then the wire width can be understood in terms of the  
edge states described in Section \ref{sec:semiclass} because they are the
propagating modes of this problem.
If the injector is narrow, i.e. it is only a few Fermi wavelength wide,
 one can assume that it excites these modes coherently. 
Therefore, as long as the distance between the injector and the collector is
smaller than the mean free path (and phase coherence lenght), 
the interference of the edge states can be important.     
Since the total wave function of the system can be written as a sum of 
all populated edge states $|k_m\rangle$, $|k_l \rangle$ at a given Fermi energy, i.e.  
$
\Psi(\mathbf{r})\sim\sum_l\mathcal{C}_l\tilde{\Phi}(y,k_l)e^{i k_l x}+
\sum_m\mathcal{C}_m\tilde{\Phi}(y,k_m)e^{i k_m x}
$ 
(here $\mathcal{C}_l$, $\mathcal{C}_m$ are normalization constants), 
interference along the confinement potential 
is determined by the phase factors $\exp(i k_{m} x)$, $\exp(i k_{l} x)$.
Here 
the wave numbers $k_{m}$, $k_{l}$ are determined by  requiring that they satisfy 
Eqs.~(\ref{eq:semi-quant-cond-1}) and (\ref{eq:semi-quant-cond-2}) respectively, 
for a given Fermi energy $E_F$ and quantum numbers $m$ and $l$.  
As in Ref.~\onlinecite{Houten_Carlo}, we  assume that the current at the collector is 
determined by the unperturbed probability density and therefore the focusing peaks are the 
results of the constructive interference of  edges states with $k_{l,m}\approx 0$ at a 
distance $x=L$ from the injector. This 
corresponds to the assumption that only electrons injected close to  perpendicular direction
can be focused onto the collector. It is convenient to introduce  the following notation: 
$\tilde{R}= R_{+}-R_{-}$, $\tilde{E}_F=E_F/\hbar\omega_c$ and we denote 
by $k_{F}^{\pm}=Q_{\pm}(E_F)/\hbar$ the radii of the two Fermi surfaces (circles) 
in the wavenumber space. Then for $k\ll k_F^{\pm}$ we have $1\gg\beta_{\pm}\approx k/k_{F}^{\pm}$ and 
expanding Eq.~(\ref{eq:semi-quant-cond-1}) in this small parameter we find that in lowest order 
\begin{subequations}
\begin{equation}
k_m=\frac{1}{4\tilde{R} \left(1+\frac{1}{\tilde{E}_F}\right)} 
\left(4\pi m-\frac{\pi}{2}[k_F^{+}R_{+}+k_F^{-}R_{-}] \right).
\label{eq:k-num-1}
\end{equation}
Similarly, expanding Eq.~(\ref{eq:semi-quant-cond-2}) we obtain
\begin{equation}
 k_l=\frac{1}{4\tilde{R}\left(1-\frac{1}{\tilde{E}_F}\right)} 
\left(2\pi (2l+1) -\frac{\pi}{2}[k_F^{+}R_{+}+k_F^{-}R_{-}] \right).
\label{eq:k-num-2}
\end{equation}
\end{subequations} 
In the semiclassical regime, where $E_F \gg \hbar\omega_c$ and hence $\tilde{E}_F \gg 1 $
we can take 
$\left(1+\frac{1}{\tilde{E}_F}\right)\approx\left(1-\frac{1}{\tilde{E}_F}\right)\approx 1$.
Therefore the phase difference $k_l L- k_m L$ at distance $L$ from the injector 
between two edge states whose wavenumbers $k_m$ and $k_l$ are given by Eq.~(\ref{eq:k-num-1}) and
Eq.~(\ref{eq:k-num-2}), respectively, reads
\begin{equation}
 k_l L- k_m L=2\pi(l-m)\frac{L}{2\tilde{R}}+\pi\frac{L}{2\tilde{R}}
\label{eq:phase-diff-2branch}
\end{equation}
and we remind that $\tilde{R}(\mathcal{B}) = \sqrt{p_{so}^2+2 m^* E_F}/{e\mathcal{B}}$. 
We see  that if for a given magnetic field e.g. $L=2\tilde{R}(\mathcal{B})$ then the phase 
difference $k_l L- k_m L$ will be an odd multiple of $\pi$. 
This means that the spinor part of the wave function 
of these two edge states, $\tilde{\Phi}(y,k_l)$ and $\tilde{\Phi}(y,k_m)$ 
[see Eq.~(\ref{eq:semi-wavefunc-spinor})] will appear with opposite 
signs in the total wave function, which may lead to a near cancellation of 
$|k_m\rangle$ and $|k_l\rangle$.

In general, whenever $L$ is an
odd multiple of $2\tilde{R}$, the phase difference will be an odd multiple of $\pi$ meaning
that there may be a near cancellation between $|k_l \rangle$ and  $|k_m \rangle$.
On the other hand, the phase difference between edge states  $|k_m \rangle$ and  $|k_{m^{'}} \rangle$,   
both belonging to the same semiclassical quantization branch given in Eq.~(\ref{eq:semi-quant-cond-1})  is 
\begin{equation}
k_m L -k_{m^{'}}L = 2\pi \frac{m-m^{'}}{\left(1+\frac{1}{4\tilde{E}}\right)} \frac{L}{2\tilde{R}}
\end{equation}
and a similar expression can be derived for the phase difference 
$k_l L -k_{l^{'}}L$ between edge states  $|k_l \rangle$ and  $|k_{l^{'}} \rangle$ 
of the other quantization branch  [Eq.~(\ref{eq:semi-quant-cond-2})]. As long as 
$\tilde{E}\gg 1$,  edge states given by the same quantization branch 
can interfere constructively at distances $L=q\, 2 \tilde{R}(\mathcal{B})$, $q=1,2,3,...$ i.e. 
regardless of whether $L$ is an even or odd multiple of the cyclotron diameter $2 \tilde{R}(\mathcal{B})$. 
By constructive interference we mean that wave function of $|k_m\rangle$ and $|k_m'\rangle$ 
have the same global sign. 
(Note however that the   $\tilde{E}\gg 1$, i.e. $E_F\gg \hbar\omega_c$ condition gives an upper
limit for the $q$ values for which this reasoning is applicable).

In contrast,  if  $L$ is an even multiple of $2\tilde{R}$ 
(e.g. $L=4\tilde{R}$) we find from Eq.~(\ref{eq:phase-diff-2branch}) that 
the phase difference   $k_l L- k_m L$  will be an integer
multiple of $2\pi$. This means  that in this case not only edge states  belonging to the same   
quantization branch but also those belonging to different quantization branches 
can interfere constructively.
Following the reasoning of Ref.~\onlinecite{Houten_Carlo} therefore 
we expect that there will be  peaks in the focusing spectra for
magnetic fields $\mathcal{B}$ where $L$ is even multiple of $2\tilde{R}(\mathcal{B})$ 
while we might not see peaks at magnetic fields corresponding to $L$ being 
odd multiple of $2\tilde{R}(\mathcal{B})$. This simple analysis then suggests that the focusing
fields are given by integer multiples of $B_{focus}=4 \sqrt{2 m^* E_{F}+p_{so}^2}/e L$.

\begin{figure}
\includegraphics[scale=0.7]{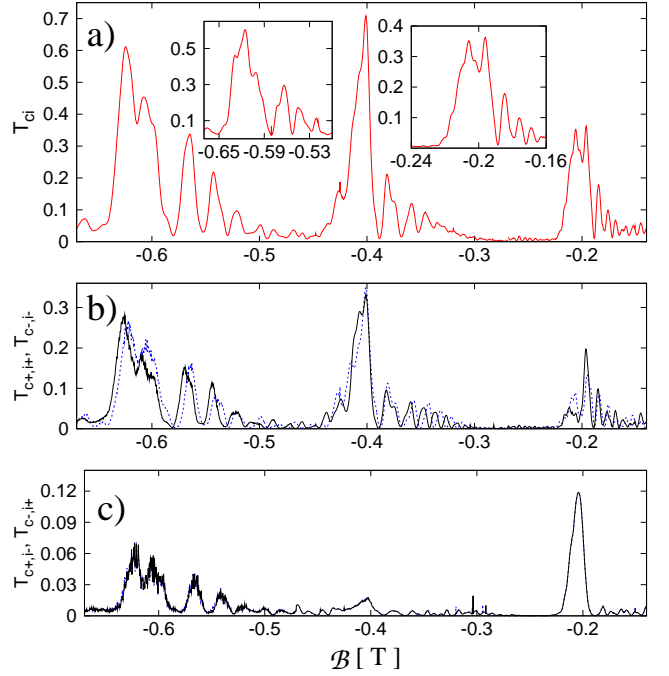}
\caption{(color online) a) The transmission from the injector to the collector as a function of the 
magnetic field. The right inset shows the close-up of the first focusing peak. The left inset shows 
that if one neglects the Zeeman interaction then the third peak is not split.
b) the partial transmissions 
$T_{c+,i+}$ (solid line) and  $T_{c-,i-}$ (dashed line). c) the partial transmissions 
$T_{c-,i+}$ (solid line) and  $T_{c+,i-}$ (dashed line) - in this scale $T_{c-,i+}$ cannot  
be distinguished from $T_{c+,i-}$. Note the different vertical scale with respect to 
Fig.~\ref{fig:focusing-main}(b).
\label{fig:focusing-main}}
\end{figure}

It is interesting to compare these predictions on the 
focusing spectra  to exact numerical transport calculations. 
Using the tight binding version\cite{reynoso-2} of the Hamiltonian (\ref{eq:gen-so}), 
the transmission probability $T_{ic}(\mathcal{B})$ between the injector and collector was calculated by 
employing the Green's function technique of Ref.~\onlinecite{sanvito}. 
The scattering region was of finite width $W$\cite{tightbind-params}  and was assumed to be perfectly ballistic 
and infinitely long (see Fig.~\ref{fig:focusing-geometry}). 
This means that the left and right ends of it act as drains which absorb any particles exiting
to the left of right.  The spin orbit coupling had a finite value in the scattering region but 
was set to zero in the injector and collector. 
To simulate the effect of finite temperatures we used a  simple energy averaging procedure in the 
calculation of the transmission curves:
$T_{ci}(\mathcal{B})=\int T_{ci}(\mathcal{B},E)\left(-\frac{\partial f_0(E)}{\partial E}\right) dE$ where 
$f_0(E)$ was the Fermi function. 
The actual results shown in Fig.~\ref{fig:focusing-main} were calculated at $T=1K$ temperature.
As it can be expected, higher temperatures tend to smear the curves while at lower ones an
additional fine structure appears. Our calculations are very similar to those 
in Ref.~\onlinecite{usaj} except that we used slightly different contacts 
(see the inset of Fig.~\ref{fig:focusing-geometry}). 
The contacts were always of a finite width (typically five-nine sites wide) and could in principle 
accommodate more than one (spin-degenerate) open channels. 

Figure.~\ref{fig:focusing-main}(a) shows the transmission $T_{ci}$ 
a function of the magnetic field $\mathcal{B}$. We used parameter values that approximately correspond  to 
the measurements of Ref.~\onlinecite{gilbertson-2} on InSb quantum wells: 
electron density $n_e=3.25*10^{15}/m^2$, effective mass $m^*=0.02\, m_e$ and Rashba coefficient 
$\alpha_R=1.4*10^{-11} {\rm eVm}$.   It has been shown\cite{nedniyom} that the effective giromagnetic factor
${\rm g}_{eff}$ of InSb is quite large and therefore the Zeeman spin splitting can be noticeable already 
at relatively weak magnetic fields. Therefore in our numerical calculation we took into account 
the Zeeman term as well and  assumed ${\rm g}_{eff}=-22$. The 
distance $L$ was $945{\rm nm}$ and both contacts were tuned to accommodate one (spin-degenerate) open channel. 
We find that for these parameters the first focusing peak is split [see the right inset of  
Fig.~\ref{fig:focusing-main}(a)]. That for strong enough 
$\alpha_R$ the first peak is split was first noticed in Ref.~\onlinecite{usaj}. 
The peak splitting  in good approximation corresponds to  
$\Delta\mathcal{B}=\frac{4 \hbar k_{so}}{e L}\approx 10\,{\rm mT}$. Comparing now the 
numerical result on the first focusing peak to our analytical prediction, we see that the dip between
the peaks is at the magnetic field value where the analytics predicts that destructive interference 
may take place for $L=2 \tilde{R}$ (corresponding to $\mathcal{B}\approx 0.2{\rm T}$) 
but that the presence of the twin peaks 
at $\mathcal{B}_-=-0.196\,{\rm T}$ and $\mathcal{B}_{+}=-0.206\,{\rm T}$  
is not explained by our approach. 
It appears that the most straightforward way to understand them is to assume spin-split 
cyclotron orbits, as in Refs.~\onlinecite{usaj,reynoso-1,zulicke}.      
Looking at the second focusing peak, one observes  that it is located at $L=4\tilde{R}$ 
(which happens for $\mathcal{B}=0.4\,{\rm T}$) quite accurately, in accordance with our edge-state based theory. 
The fact that its amplitude is significantly larger than the 
amplitude of the first peaks seems to corroborate the theoretical prediction that in this case 
edge states belonging to different quantization branches 
[Eq.~(\ref{eq:semi-quant-cond-1}) and Eq.~(\ref{eq:semi-quant-cond-2})] 
can constructively interfere with each other. Whether this enhancement of the amplitude
could be observed  in an actual experimental situation, however, 
depends on how specular the reflection at the confinement potential is. 
(Note that in the classical picture the second focusing peak correspond to trajectories which 
bounce off the boundary between the injector and collector once, see the dashed line in 
Fig.~\ref{fig:focusing-geometry}.) A small amount of diffuse scattering
at the boundary may render the observation of this enhancement  difficult. Finally, we find that  close to 
$L=6\tilde{R}$ ($\mathcal{B}=0.597{\rm T}$), where our calculations  predict 
that the wave functions of the edge states may cancel,  the amplitude of
the transmission  is indeed  small, but the focusing peak at a slightly higher magnetic field is again
not captured by our calculations. It seems that peaks which appear when $L$ is odd multiple
of $2\tilde{R}$ can not be described with the presented theoretical approach. 
The  splitting of the third peak close to 
$\mathcal{B}=-0.62{\rm T}$, reminiscent of the splitting of the first one, 
is due to the Zeeman interaction and not to the SOI. As we mentioned, because of the large 
${\rm g}_{eff}$ the Zeeman energy can be important at smaller magnetic fields than in e.g. GaAs. 
This is illustrated in the left inset of  Fig.~\ref{fig:focusing-main}(a) where we show the calculation for 
the third peak but without taking into account the Zeeman term in the Hamiltonian. One can see that 
for the assumed strength of $\alpha_R=1.4*10^{-11}{\rm eVm}$ the third peak is not split in this case.  

In Fig.~\ref{fig:focusing-main}(b) and (c) we show the partial transmissions assuming spin polarized 
injection/detection. Thus,  e.g. $T_{c+,i-}$  refers to the transmission probability of electrons being 
injected in spin $-1$ eigenstate and collected in $+1$.   
In contrast to Ref.~\onlinecite{usaj} we chose as  spin quantization axis 
the $\hat{y}$ direction (for the definition of the coordinates, see Fig.~\ref{fig:focusing-geometry}).
The motivation to choose this axis comes from Ref.~\onlinecite{grigoryan} where it was shown that
the average spin of the edge states (at least the low energy ones) pointed mainly in the 
direction perpendicular to the confinement potential, i.e. along the ${y}$ axis in our case.  
Comparing Fig.~\ref{fig:focusing-main}(b) and (c) one can observe that 
except for the first focusing peak, the spin-flip transmissions $T_{c+,i-}$, $T_{c-,i+}$ are always 
significantly smaller than $T_{c+,i+}$, $T_{c-,i-}$. We also performed calculations (not shown here) 
where the  injected electrons were polarized in the $\hat{x}$ direction, as in Ref.~\onlinecite{usaj}, and we 
found that $T_{c+,i-}$, $T_{c-,i+}$ were, apart from the vicinity of the first peak, 
usually smaller in the case of  $\hat{y}$ polarized injection than for $\hat{x}$ polarized one. 
Further investigation of the average polarization of the spin of the edge states in the semiclassical limit and its 
effect on the partial transmissions in a focusing setup is left to  a future work.

\section{Summary}
\label{sec:summary}

In summary, we studied the role of edge states in  transverse electron focusing  setup 
for strong spin-orbit coupling. As a first step, employing a semiclassical approach we derived
a  good approximation for the dispersion relation of the edge states and briefly compared our 
results to other approximation methods that can be found in the literature.
We then studied the interference of the edge states as this is expected to have ramifications 
on the focusing spectrum. Comparison of our theoretical results with  numerical transport calculations 
suggests that certain properties of 
the focusing spectrum can be quite well understood in terms of the interference of the edge states.  
Nevertheless, the presented semiclassical approach can not capture all the important characteristics
of the transport calculations. Finally, we studied numerically the electron focusing when spin polarized
injection was used.

\emph{Acknowledgements}: A. K. was supported by EPSRC.


\begin{thebibliography}{10}
\bibitem{tightbind-params}
The tight binding parameters we used in the numerical computations shwon in Fig.~\ref{fig:focusing-main} 
were as follows: in terms of the lattice constant $a_0$, the scattering region was $W/a_0=120$ wide, the 
distance between the injector and collector was  $L/a_0=135$, 
the width of the probes was $W_{I}/a_0=W_{C}/a_0=4$ and the Rashba coupling constant was 
$\alpha_R/(2 a_0\gamma)=0.027$, where the hopping $\gamma$ was set to unity.
\end{thebibliography}
\end{document}